\documentstyle[12pt,epsf]{article}

\input{tcilatex}

\begin{document}

\begin{titlepage}

{\noindent \Large \bf On the measure of nonclassicality of field states}\\
 
\vspace{0.3cm} 
{\noindent \large\em J.M.C. Malbouisson${}^{(1)}$ and B. Baseia${}^{(2)}$}\\
 
\noindent ${}^{(1)}$ {Instituto de F\'{\i}sica, Universidade Federal da Bahia, 
40210-340, Salvador, BA, Brazil}\\ 
\noindent ${}^{(2)}$ {Instituto de F\'{\i}sica, Universidade Federal da Goi\'as, 
74001-970, Goi\^ania, GO, Brazil}\\
\vspace{0.6cm}  
\noindent E-mail: jmalboui@ufba.br, basilio@fis.ufg.br\\ 
\vspace{0.1cm} 
\noindent PACS Ref: 03.65.Bz, 42.50.Dv, 42.65.Ky

\subsection*{Abstract}   
The degree of nonclassicality of states 
of a field mode is analysed considering both phase-space and 
distance-type measures of nonclassicality. By working out some general examples, 
it is shown explicitly that the phase-space measure is rather sensitive
to superposition of states, with finite superpositions  
possessing maximum nonclassical depth (the highest degree of
nonclassicality) irrespective to the nature of
the component states. Mixed states are also discussed and examples with
nonclassical depth varying between the minimum and the maximum allowed
values are exhibited. For pure Gaussian states, it is demonstrated that 
distance-type measures based on the Hilbert-Schmidt metric are equivalent 
to the phase-space measure. Analyzing some examples, it is shown that 
distance-type measures are efficient to quantify the degree of nonclassicality 
of non-Gaussian pure states.

\end{titlepage} 
\newpage \baselineskip.37in

\section{Introduction}

Since the first observations of nonclassical properties of electromagnetic
field states,\ like sub-Poissonian statistics \cite{subP} and squeezing \cite
{squeez}, theoretical interest has been growing in the search of a way to
measure the degree of nonclassicality of a given state. Generally, one can
classify a state as nonclassical if its Glauber-Sudarsham $P$-function \cite
{GS} has negative values or if it is more singular than a delta function.
However, the nonclassical properties do not occur simultaneously for all the
states; for example, squeezed-vacuum states are super-Poissonian while the
odd coherent states are sub-Poissonian but do not present quadrature
squeezing, \noindent and so the question how much nonclassical a given state
is seems to be appropriated.

Several proposals have been made to quantify how much nonclassical a state
can be. Restricted to the photon statistics, Mandel \cite{Mandel} introduced
a parameter ($q=\left\langle \hat{n}^{2}\right\rangle /\left\langle \hat{n} 
\right\rangle -\left\langle \hat{n}\right\rangle -1$) quantifying the
departure from the Poissonian behavior of a coherent (quasi-classical)
state; such a measure, however, does not contemplate other nonclassical
properties of \ field states. Hillery \cite{Hillery}, defined the
nonclassical distance of a state as the trace norm of the difference between
the density operator of the state and that of the nearest classical state.
In practical calculations, however, the determination of the nearest
classical state is rather difficult. Other more operational measures of
nonclassicality, following this trend, have been introduced by Dodonov {\it  
et al.} \cite{Dodonov}, using the Hilbert-Schmidt distance between density
operators and, more recently, by Marian {\it et al.} \cite{Marian},
employing the Bures-Uhlmann definition of distance between states.
Distinctly, inspired in the Cahill-Glauber representation \cite{CG}, Lee 
\cite{Lee1} introduced the $R$-function as a (real) $\tau $-parametrized
Gaussian convolution of the $P$-function and defined the nonclassical depth
of a state as the minimum value of $\tau $ for which the $R$-function
becomes a nonnegative definite function, thus acceptable as a classical
distribution function. A similar phase-space measure of nonclassicality was
also considered by L\"{u}tkenhaus and Barnett \cite{LB}. Nonclassical
measure of nonclassical properties \cite{Kim} and an observable criterion to
distinguish nonclassical states \cite{Vogel} have also been discussed
recently.

This article focuses initially on the phase-space measure of nonclassicality
of one-mode field states as firstly introduced by Lee \cite{Lee1}. After a
brief review of the phase-space criterium, by analyzing some general cases,
it is explicitly shown that this measure is rather sensitive to
superposition of states, with finite superpositions of states having the
highest degree of nonclassicality (maximum nonclassical depth) irrespective
to the nature of the component states. Next, the nonclassical depth of mixed
states is considered and examples with nonclassical depth varying between
the minimum and the maximum permitted values are presented. The paper
proceeds with the analysis of distance-type measures. It is demonstrated
that distance-type measures based on the Hilbert-Schmidt metric are
equivalent to the phase-space measure for pure Gaussian states (squeezed
states). Then, it is shown that they provide a way of quantifying the degree
of nonclassicality for non-Gaussian ones. Finally, some conclusions are
drawn.

\section{Phase-space measure of nonclassicality}

By noticing that Gaussian convolutions of the $P$-function, as the Wigner ($ 
W $) and the Husimi ($Q$) functions, are regular functions of a complex
variable $z$ and inspired in the Cahill-Glauber representation \cite{CG},
Lee \cite{Lee1} introduced a continuous family of representations of a field
state depending on one parameter ($\tau \geq 0$) as the set of Gaussian
convolutions of the $P$-function 
\begin{equation}
R(z;\tau )=\frac{1}{\pi \tau }\int d^{2}w\,\exp \left( -\frac{1}{\tau } 
|z-w|^{2}\right) P(w)\,;  \label{Rfunction}
\end{equation}
$\tau =\frac{1}{2}$ and $1$ correspond to the $W$- and the $Q$-functions
respectively, while one recovers the $P$-function in the limit $\tau
\rightarrow 0$. The fact that the $W$-function is regular but usually
possesses negative values and that the $Q$-function is always a non-negative
definite function leads to the definition of the nonclassical depth of a
given state as the minimum value of $\tau $ ($\tau _{{\rm m}}$) for which
the corresponding $R$-function is non-negative in the whole $z$-plane, thus
becoming acceptable as a classical distribution function. From this
definition, it follows that $0\leq \tau _{{\rm m}}$ $\leq 1$, examples of
limiting cases being coherent and thermal states, with $\tau _{{\rm m}}$ $=0$ 
, and number states which have maximum nonclassical depth ($\tau _{{\rm m}}$ 
$=1$). As non extreme examples, squeezed states possess $\tau _{{\rm m}}$
ranging from $0$ to $\frac{1}{2}$ as the squeezing parameter ($e^{|\zeta |}$ 
) varies from $1$ to $\infty $. Not only this criterium is well-defined
mathematically but also it indicates how to calculate the degree of
nonclassicality (the nonclassical depth) of a given state of the field.

If one works in the number basis, with matrix elements of the density
operator of a given state denoted by $\rho (n,m)=\left\langle n|\hat{\rho} 
|m\right\rangle $, the $R$-functions are given by \cite{Lee3} 
\begin{equation}
R(z;\tau )=\frac{1}{\pi \tau }e^{-|z|^{2}/\tau }\sum_{n,m=0}^{\infty }\rho
(n,m)e^{i(m-n)\theta }A_{n,m}  \label{Rnum}
\end{equation}
where $z=|z|e^{i\theta }$, 
\begin{equation}
A_{n,m}=\left( \frac{m!}{n!}\right) ^{1/2}\frac{(\tau -1)^{m}}{\tau ^{n}} 
\,|z|^{n-m}L_{m}^{n-m}\left( \frac{|z|^{2}}{\tau (\tau -1)}\right)
\;\;,\;\;n\leq m\;,  \label{Anm}
\end{equation}
$L_{m}^{n-m}$ denoting the generalized Laguerre polynomial, and with $ 
A_{n>m}=A_{m,n}$. This expression is very convenient to calculate the $R$
-functions of finite superpositions of number states. On the other hand, in
the coherent basis, the $R$-functions can be written as \cite{Lee1} 
\begin{eqnarray}
R(z;\tau ) &=&\frac{1}{\pi (1-\tau )}\exp \left( \frac{|z|^{2}}{1-\tau } 
\right) \int d^{2}\beta \left\langle -\beta \left| \hat{\rho}\right| \beta
\right\rangle  \nonumber \\
&&\;\;\;\;\;\;\;\;\;\;\;\times \exp \left( -\frac{1}{1-\tau }\left[ (2\tau
-1)\left| \beta \right| ^{2}+(z^{\ast }\beta -z\beta ^{\ast })\right]
\right) \;.  \label{Rcoh}
\end{eqnarray}
For a general finite superposition of coherent states, 
\begin{equation}
\left| \psi \right\rangle =\sum_{i=1}^{N}c_{i}\left| \alpha
_{i}\right\rangle \;,  \label{phisi}
\end{equation}
performing some Gaussian integrals, the Eq. (\ref{Rcoh}) reduces to 
\begin{eqnarray}
R(z;\tau ) &=&\frac{1}{\pi \tau }\sum_{i,j=1}^{N}c_{i}c_{j}^{\ast }\exp
\left( -\frac{1}{2}\left[ |\alpha _{i}|^{2}+|\alpha _{j}|^{2}-2\alpha
_{i}\alpha _{j}^{\ast }\right] \right)  \nonumber \\
&&\;\;\;\;\;\;\;\;\;\times \exp \left( -\frac{1}{\tau }(z-\alpha
_{i})(z^{\ast }-\alpha _{j}^{\ast })\right) \;.  \label{Rcoherent}
\end{eqnarray}
This formula is useful to calculate the nonclassical depth of superpositions
of coherent states.

\section{Nonclassical depth of superposed states}

Since the $R$-functions are certainly regular functions of $z$ for $\tau
\geq 1/2$, a general procedure to find the nonclassical depth of a state ($ 
\tau _{{\rm m}}$ as defined before) consists in determining the minimum
value of the function $R(z;\tau )$ for a sequence of values of $\tau $ and
then search for the smallest value of $\tau \in (0,1]$ for which the minimum
value of $R(z;\tau _{{\rm m}})$ vanishes. In many cases, where one suspects
that $\tau _{{\rm m}}$ $=1$ (maximum nonclassical depth), one may
investigate how the value of $R(z;\tau )$ at a chosen point behaves as $\tau 
$ is increased inside the interval $(0,1]$. Next, the nonclassical depth of
some examples of superposed states of the electromagnetic field are
calculated.

\subsection{General superposition of the vacuum and the $n$-photons state}

As a first example, consider superpositions of the vacuum $\left|
0\right\rangle $, a state with minimum degree of nonclassicality ($\tau _{ 
{\rm m}}$ $=0$), with the number state $\left| n\right\rangle $, which has
maximum nonclassical depth ($\tau _{{\rm m}}$ $=1$). Take the set of
normalized states 
\begin{equation}
\left| \psi (\xi ;\phi )\right\rangle _{0,n}=\sqrt{\xi }\,\left|
0\right\rangle +\sqrt{1-\xi }\,e^{i\phi }\,\left| n\right\rangle \;,
\label{S01}
\end{equation}
where $0\leq \xi \leq 1$ and $\phi $ is an arbitrary relative phase. These
states continuously interpolate from $\left| 0\right\rangle $ to $\left|
n\right\rangle $, so one wonders what happens to the nonclassical depth of
such states, as the parameter $\xi $ varies from $0$ to $1$. From Eqs. (\ref
{Rnum}) and (\ref{Anm}), it follows that the $R$-function corresponding to
the state (\ref{S01}) is given by 
\begin{eqnarray}
R_{\left| \psi \right\rangle _{0,n}}(z;\tau ) &=&\frac{1}{\pi \tau } 
e^{-|z|^{2}/\tau }\left[ \xi L_{0}\left( \frac{|z|^{2}}{\tau (\tau -1)} 
\right) +(1-\xi )\left( \frac{\tau -1}{\tau }\right) ^{n}L_{n}\left( \frac{ 
|z|^{2}}{\tau (\tau -1)}\right) \right.  \nonumber \\
&&\;\;\;\;\;\;\;\;\;\;\;\;\;\;\;\;\left. +2\sqrt{\xi (1-\xi )}\left( \frac{ 
|z|}{\tau }\right) ^{n}L_{0}^{n}\left( \frac{|z|^{2}}{\tau (\tau -1)}\right)
\cos \left( n\theta -\phi \right) \right] \;,  \label{Rfunc0n}
\end{eqnarray}
and one can analyze its minimum value for various cases.

For the simplest one, $n=1$, using that $L_{0}(x)=1$, $L_{1}(x)=1-x$ and $ 
L_{0}^{1}(x)=1$, Eq. (\ref{Rfunc0n}) reduces to

\begin{equation}
R_{\left| \psi \right\rangle _{0,1}}(z;\tau )=\frac{1}{\pi \tau } 
e^{-|z|^{2}/\tau }\left[ \frac{(1-\xi )}{\tau ^{2}}|z|^{2}+\frac{2\sqrt{\xi
(1-\xi )}}{\tau }\cos (\theta -\phi )|z|+\frac{\xi +\tau -1}{\tau }\right]
\;.
\end{equation}
One sees that the sign of the minimum value of $R_{\left| \psi \right\rangle
_{0,1}}$ is dictated by the term inside the square bracket above. If one
chooses the worse circumstance as far as the positiveness of the function $ 
R_{\left| \psi \right\rangle _{0,1}}$ is concerned, $\cos (\theta -\phi )=-1$
(corresponding to look at the axis in the $z$-plane defined by $\theta =\phi
+\pi $), one sees that the minimum of the quadratic function of $|z|$ in the
square bracket above is attained at $|z|=\tau \sqrt{\xi /(1-\xi )}$ and has
value $(1-\xi )(\tau -1)/\tau $, which is negative for all $0<\tau <1$ for
any $0\leq \xi <1$, being zero only when $\tau =1$. Therefore, the states $ 
\left| \psi (\xi ;\phi )\right\rangle _{0,1}$ possess $\tau _{{\rm m}}$ $=1$
for all $0\leq \xi <1$ and such states are as nonclassical as possible. It
is impressive the fact that, no matter how close the value of $\xi $ is to $ 
1 $, independently of the relative phase $\phi $, the state $\left| \psi
(\xi ;\phi )\right\rangle _{0,1}$ is as much nonclassical as it is a number
state, in particular the state $\left| 1\right\rangle $. In contrast, the
Mandel parameter for the superposition (\ref{S01}) is given by $q_{\left|
\psi \right\rangle _{0,n}}=-1+n\xi $ and so, for $n=1$, it varies
continuously from $-1$ to $0$ as $\xi $ goes from $0$ to $1$. Analyzing the
minimum of the function (\ref{Rfunc0n}), one can verify that $\tau _{{\rm m} 
} $ $=1$ is maintained for the cases $n>1$. One is then led to speculate
that the phase-space measure of nonclassicality is rather sensitive to
superposition of states and that maximum nonclassical depth is more a rule
than an exception. To check this assertion, the superposition of two
quasi-classical states are investigated next.

\subsection{General superposition of opposite coherent states}

Consider a general superposition of opposite coherent states ($\alpha $ real
for simplicity) given by 
\begin{equation}
\left| \psi (\xi ,\phi )\right\rangle ={\cal N}\left( \sqrt{\xi }\left|
-\alpha \right\rangle +\sqrt{1-\xi }e^{i\phi }\left| \alpha \right\rangle
\right) \;,
\end{equation}
for which ${\cal N}=\left( 1+2\sqrt{\xi (1-\xi )}\cos \phi \,e^{-2\alpha
^{2}}\right) ^{-1/2}$; such state interpolates between the coherent states $ 
\left| \alpha \right\rangle $ and $\left| -\alpha \right\rangle $ as $\xi $
varies from $0$ to $1$. For $|\alpha |\gg 1$, $\left\langle -\alpha |\alpha
\right\rangle =\exp (-2\alpha ^{2})\approx 0$, ${\cal N}\cong 1$ and such
states correspond to mesoscopic superpositions of quasi-classical states
known as Schr\"{o}dinger-cat states of the field. In the particular case
where $\xi =1/2$ and $\phi =0$ ($\phi =\pi $), this superposition is called
even (odd) coherent state and possesses well-defined parity. Since both
component states have minimal nonclassical depth, $\tau _{{\rm m}}$ $=0$, it
is worth looking at the nonclassical depth of the superposition. Using
expression (\ref{Rcoherent}), one has 
\begin{equation}
R_{\left| \psi \right\rangle }(z;\tau )=\frac{1}{\pi \tau }\left( e^{\alpha
^{2}}+2\sqrt{\xi (1-\xi )}\cos \phi \,e^{-\alpha ^{2}}\right) ^{-1}\exp
\left( -\frac{1}{\tau }\left( x^{2}+y^{2}\right) \right) F(x,y;\tau ;\xi
;\phi )  \label{Rcat}
\end{equation}
where $z=x+iy$ and 
\begin{eqnarray}
F(x,y;\tau ;\xi ;\phi ) &=&\exp \left( -\frac{(1-\tau )}{\tau }\alpha
^{2}\right) \left[ \exp \left( \frac{2\alpha x}{\tau }\right) -2\xi \sinh
\left( \frac{2\alpha x}{\tau }\right) \right]  \nonumber \\
&&\;\;\;\;\;\;\;\;+2\sqrt{\xi (1-\xi )}\exp \left( \frac{(1-\tau )}{\tau } 
\alpha ^{2}\right) \cos \left( \frac{2\alpha y}{\tau }-\phi \right) \;.
\label{Fcat}
\end{eqnarray}

The function $F$, which determines the sign of\ the R-function (\ref{Rcat}),
possesses minima located at points $x=\tau \left[ \ln \xi -\ln (1-\xi ) 
\right] /4\alpha $ and $y=\left[ (2n+1)\pi +\phi \right] \tau /2\alpha $ ,
where $n$ is an integer. The minimum values attained are all identical and
given by $F_{\min }=-4\sqrt{\xi (1-\xi )}\sinh \left[ \alpha ^{2}(1-\tau
)/\tau \right] $, which is negative for all $0<\tau <1$, for $0<\xi <1$ and
independently of $\phi $, no matter the value of $\alpha $. Therefore, one
concludes that superpositions of opposite coherent states possess maximum
nonclassical depth ($\tau _{{\rm m}}$ $=1$), irrespective of the value of $ 
\alpha $, notwithstanding the fact that components of these superpositions
are as classical as possible, having $\tau _{{\rm m}}$ $=0$. It is amazing
that a superposition of two quasi-classical states (with $\tau _{{\rm m}}$ $ 
=0$), no matter how further apart they are, is as nonclassical as possible,
with the same nonclassical depth ($\tau _{{\rm m}}$ $=1$) as that for a
number state. It seems though that just the fact of considering a
superposition of states, a concept lying at the heart of Quantum Mechanics,
is enough to lead to maximum degree of nonclassicality in the sense of the
phase-space measure.

Many other examples can be treated, corroborating with the point discussed
above: superpositions of coherent and number states, of squeezed and number
states, of coherent and squeezed states and so on. In all such cases, one
finds maximum nonclassical depth. Also, higher-generation Schr\"{o} 
dinger-cat states of a field mode \cite{MB} possess maximum degree of
nonclassicality in the context of the phase-space measure. As another
important example, it has been shown that all Pegg-Barnett phase-states \cite
{PB} have the greatest nonclassical depth ($\tau _{{\rm m}}$ $=1$),
irrespective to the dimension of the truncated Hilbert space they are
defined \cite{MBGB,Jorge}. Thus the phase-space criterium of nonclassicality
for pure states of the field looks like an ``all-or-nothing''\ measure, with 
$\tau _{{\rm m}}$ being either $0$ or $1$, with the exception of squeezed
states. Also, it is clear that no direct relation between the phase-space
measure of nonclassicality and the presence of some specified nonclassical
property can exist; for example, a state with $\tau _{{\rm m}}$ $=1$ can be
super-Poissonian (as phase states) or can present no quadrature squeezing,
as happens with number states.

It is worth comment on the equivalence between the phase-space measure of
nonclassicality as stated by Lee \cite{Lee1} and by L\"{u}tkenhaus and
Barnett \cite{LB}. In the later, maximum nonclassicality is signalized by
the existence of zeros in the $Q$-function while Lee's criterium for maximum
nonclassical depth can be stated by saying that the $R$ -functions for all $ 
\tau <1$ possess negative regions. The fact that $R(z;\tau )$, for all $z$,
are continuous functions of the parameter $\tau $ ensures the mentioned
equivalence. For the simple examples analyzed here, one can easily follow
the disappearance of the negative regions of the $R$ -functions as $\tau $
approaches $1$. Actually, by analyzing the zeros of the Husimi $Q$-function,
it can be proved that all quantum states which are not Gaussian have maximum
degree of nonclassicality in this phase-space measure \cite{LB}. The
insensitivity of the phase-space measure to distinguish between pure,
non-Gaussian, states raises the question whether other type of measure may
be more efficient to do so. As it will be shown in Section 5, measures based
on notions of distance between states are able to quantify in a broader
sense the nonclassicality of pure states. But, before presenting this
analysis, the phase-space measure of nonclassicality of mixed states is
considered.

\section{Nonclassical depth of mixed states}

Now, turn the attention to mixed states. Clearly, since the phase-space
representations are linear in the density operator, if one considers a
mixture of states with all parcels possessing $\tau _{{\rm m}}$ $=0$, the
nonclassical depth is maintained minimal. In the same way, mixtures of
states having maximum nonclassical depth possess $\tau _{{\rm m}}$ $=1$. On
the other hand, Lee \cite{Lee3} showed that any state (pure or not) for
which $\rho (0,0)=\left\langle 0|\hat{\rho}|0\right\rangle =0$ possesses
maximum nonclassical depth. An explicit example is the mixed state described
in Ref. \cite{BDM}, which interpolates between a number and a chaotic
(thermal) state. So, only mixtures containing a non-null vacuum part need to
be analyzed and, for simplicity, take general mixtures of $\left|
0\right\rangle $ and a number state $\left| n\right\rangle $, specified by 
\begin{equation}
\hat{\rho}=\xi \,\left| 0\right\rangle \left\langle 0\right| +(1-\xi )\left|
n\right\rangle \left\langle n\right| \;;  \label{mix}
\end{equation}
such mixed states also interpolates between $\left| 0\right\rangle $ and $ 
\left| n\right\rangle $, but in a distinct fashion from that of the state ( 
\ref{S01}). The $R$-functions, in this case, are given by 
\begin{equation}
R_{\hat{\rho}}(z;\tau ;\xi )=\frac{1}{\pi \tau }e^{-|z|^{2}/\tau }\left[ \xi
+(1-\xi )\left( \frac{1-\tau }{\tau }\right) ^{n}(-1)^{n}L_{n}\left( \frac{ 
|z|^{2}}{\tau (1-\tau )}\right) \right] \;,  \label{Rmix}
\end{equation}
and the investigation of the minimum value of the function within the square
bracket leads to the nonclassical depth.

In the extremes, $\xi =0$ and $\xi =1$, the nonclassical depth are $1$ and $ 
0 $, respectively, while for $0<\xi <1$, it is found that $\tau _{{\rm m} 
}^{(n)}$ $(\xi )$ varies continuously between $1$ and $0$. For small $n$, $ 
\tau _{{\rm m}}^{(n)}$ can be calculated analytically and one finds, for
example, $\tau _{{\rm m}}^{(1)}(\xi )=1-\xi $ and $\tau _{{\rm m}}^{(2)}(\xi
)=\left( \sqrt{\xi (1-\xi )}-1+\xi \right) /\left( 2\xi -1\right) $.
Fig.\thinspace 1 illustrates the behavior of $\tau _{{\rm m}}^{(n)}(\xi )$
for some values of $n$. One sees that, as $n$ increases, the nonclassical
depth grows for $\frac{1}{2}<\xi <1$, showing that the nonclassicality of
the number state dominates the scenario when $n$ is large even when the
weight of the vacuum part in the mixture is bigger than that of $\left|
n\right\rangle $. There is a competition between the positiveness of the $R$ 
-function of the vacuum (for all $0\leq \tau \leq 1$) and the existence of
negative regions in the $R$ -function of the number state $\left|
n\right\rangle $ when $0<\tau <1$ but, in a sense, the number state is
favored.

As far as the nonclassical properties of the states (\ref{mix}) are
concerned, it is interesting to notice that the Mandel factor is given by $ 
q(\xi ,n)=\xi n-1$ (the same expression as that for the superposition (\ref
{S01}), as expected) and one sees that super-Poissonian statistics occurs
for all $n\geq 2$ if $\frac{1}{n}<\xi <1$; for $\xi >\frac{1}{2}$, the
super-Poissonian character grows as $n$ increases, even though the states
become more nonclassical within the phase-space measure. Furthermore, these
states do not present quadrature squeezing in any circumstance. This shows
that no direct relation between the nonclassical depth and these
nonclassical properties exists. Also, there is no relation between the
nonclassical depth and the degree of impurity $D={\rm Tr}\left( \hat{\rho}- 
\hat{\rho}^{2}\right) $, measuring the departure from the idempotent
property; states (\ref{mix}) have $D=2\xi (1-\xi )$ independently of $n$,
while $\tau _{{\rm m}}^{(n)}(\xi )$ grows as $n$ increases.

Other situations can be analyzed (as mixtures of coherent and number states,
of squeezed and number states, and so on) suggesting that a mixture of two
states has nonclassical depth varying continuously in the interval defined
by the nonclassical depth of the added states. But there are exceptions, for
example a mixture of a state with $\tau _{{\rm m}}$ $=0$ (e.g. a coherent
state) and a squeezed state. This is because the $R$-functions of a squeezed
state $\left| \zeta \right\rangle $ becomes singular when $\tau <\tau _{{\rm  
m}}^{(\zeta )}$ \cite{Lee1} and the positiveness of the $R$-functions of
states with $\tau _{{\rm m}}$ $=0$ can not compensate this singular
behavior. In such cases, the nonclassical depth of the mixture is equal to $ 
\tau _{{\rm m}}^{(\zeta )}$. For the same reason, a mixture of two squeezed
states has nonclassical depth equal to the largest value of $\tau _{{\rm m}}$
of its parts.\ 

\section{Distance-type measures of nonclassicality}

There are other measures of nonclassicality defined as functions of the
distance between the state and a conventionalized set of all classical
states of the field mode. Such kind of measure was firstly introduced by
Hillery \cite{Hillery}, with the degree of nonclassicality being given by
the trace norm of the difference between density operators of the state and
of its nearest classical state. The determination of the nearest classical
state, however, is rather difficult to be implemented. A more operational
way to introduce distance-type measures of nonclassicality consists in to
elect a subset of the space of the density operators as the set of most
classical states, choosing a well-defined distance function in it, and to
define the degree of nonclassicality as the minimum value of a monotonically
increasing function of the distance between the state and a representative
(arbitrary) nonclassical state.

Some definitions of distance between states have been used in Quantum
Optics, important examples being the Hilbert-Schmidt distance, 
\begin{equation}
d^{{\rm HS}}(\hat{\rho},\hat{\sigma})=\sqrt{{\rm Tr}\left\{ \left( \hat{\rho} 
-\hat{\sigma}\right) ^{2}\right\} }\;,\;\;\;\;\;\;\;\;\;\;  \label{HSdist}
\end{equation}
and the Bures-Uhlmann distance, 
\begin{equation}
d^{{\rm BU}}(\hat{\rho},\hat{\sigma})=\sqrt{2-2{\rm Tr}\left\{ \left( \sqrt{ 
\hat{\rho}}\,\hat{\sigma}\sqrt{\hat{\rho}}\right) ^{1/2}\right\} } 
\;.\;\;\;\;\;\;\;\;\;  \label{BUdist}
\end{equation}
These metrics have been employed to define distance-type measures of
nonclassicality by Dodonov {\it et al.} \cite{Dodonov} and by Marian {\it et
al.} \cite{Marian}, respectively. For pure states, $\left| \psi
\right\rangle $ and $\left| \varphi \right\rangle $, these distances are
expressed as functions of the quantum-mechanical transition probability
between the states, $\left| \left\langle \psi |\varphi \right\rangle \right|
^{2}\,$; the Hilbert-Schmidt distance reduces to 
\begin{equation}
d^{{\rm HS}}(\left| \psi \right\rangle ,\left| \varphi \right\rangle )=\sqrt{ 
2-2\left| \left\langle \psi |\varphi \right\rangle \right| ^{2}}\;,
\label{HSdistpu}
\end{equation}
while the Bures-Uhlmann distance becomes 
\begin{equation}
d^{{\rm BU}}(\left| \psi \right\rangle ,\left| \varphi \right\rangle )=\sqrt{ 
2-2\left| \left\langle \psi |\varphi \right\rangle \right| }\;.
\label{BUdistpu}
\end{equation}

In dealing with pure states, the natural choice as the set of most
nonclassical states consists of the set of all coherent states, $\left\{
\left| \beta \right\rangle \right\} $. Since the minimum of any
monotonically increasing function of the distance between the state and the
set of most nonclassical ones can be used as a measure of nonclassicality,
here, 
\begin{equation}
d_{{\rm m}}=\min_{\{\left| \beta \right\rangle \}}\left[ 1-\left|
\left\langle \psi |\beta \right\rangle \right| ^{2}\right] =1-\pi
\max_{\{\beta \in {\bf C}\}}Q_{\left| \psi \right\rangle }(\beta )\;,
\label{DTmeasu}
\end{equation}
where $Q_{\left| \psi \right\rangle }(\beta )$ is the Husimi $Q$-function
corresponding to $\left| \psi \right\rangle $, will be used to quantify the
degree of nonclassicality of the pure state $\left| \psi \right\rangle $.
This definition slightly differs from the measures of nonclassicality
introduced in Refs. \cite{Dodonov} and \cite{Marian}; it just avoids
unnecessary square roots and it is normalized such that $0\leq d_{{\rm m}}<1$ 
, making easy the comparison with the nonclassical depth $\tau _{{\rm m}}$.
Naturally, the minimal value ($d_{{\rm m}}=0$) occurs for coherent states.
For a number state $\left| n\right\rangle $ one has $d_{{\rm m} 
}^{(n)}=1-n^{n}e^{-n}/n!\,$, so that all number states have different
degrees of nonclassicality, the upper limit value been reached only when $ 
n\rightarrow \infty $; this feature does not happens within the context of
the phase-space measure. Notice that, while the nonclassical depth ($\tau _{ 
{\rm m}}$) is obtained from the minimum of the $R$-function, the most
nonclassical states having zeros in the corresponding $Q$-functions, $d_{ 
{\rm m}}$ is determined from the maximum value of the Husimi $Q$-function.

These measures of nonclassicality, however, are equivalent to each other if
one restricts the analysis to the set of Gaussian pure states. Indeed, for a
Stoler squeezed state 
\begin{equation}
\left| \alpha ,\zeta \right\rangle =\hat{D}(\alpha )\hat{S}(\zeta )\left|
0\right\rangle \;,  \label{squezsta}
\end{equation}
with $\hat{D}(\alpha )=\exp (\alpha \hat{a}^{\dagger }-\alpha ^{\ast }\hat{a} 
)$ and $\hat{S}(\zeta )=\exp (\frac{1}{2}\zeta \hat{a}^{\dagger 2}-\frac{1}{2 
}\zeta ^{\ast }\hat{a}^{2})$, the $Q$-function is given by (setting $\zeta
=re^{i\theta }$) 
\begin{eqnarray}
Q_{\left| \alpha ,\zeta \right\rangle }(\beta ) &=&\frac{{\rm sech}\,r}{\pi } 
\exp \left[ -(1-\tanh r\cos \theta )x^{2}\right.  \nonumber \\
&&\left. -(1+\tanh r\cos \theta )y^{2}+2\tanh r\sin \theta \,xy\right] \;,
\label{Qsqueez}
\end{eqnarray}
where $x=\func{Re}(\beta -\alpha )$ and $y=\func{Im}(\beta -\alpha )$. The
maximum value of this function is equal to $\pi ^{-1}{\rm sech}\,r$ and is
reached when $\beta =\alpha $. Therefore, for a squeezed state, 
\begin{equation}
d_{{\rm m}}^{(\zeta )}=1-{\rm sech}\,r\;.  \label{dmsqueez}
\end{equation}
In terms of the nonclassical depth for a squeezed state \cite{Lee1}, 
\begin{equation}
\tau _{{\rm m}}^{(\zeta )}=\frac{\tanh r}{1+\tanh r}\;,  \label{tmsquueez}
\end{equation}
this quantity can be written as 
\begin{equation}
d_{{\rm m}}^{(\zeta )}=1-\sqrt{\frac{1-2\tau _{{\rm m}}^{(\zeta )}}{\left(
1-\tau _{{\rm m}}^{(\zeta )}\right) ^{2}}}\;,  \label{dmtmsq}
\end{equation}
that is, $d_{{\rm m}}^{(\zeta )}$ is a bijective function of $\tau _{{\rm m} 
}^{(\zeta )}$, which proves the equivalence between these two measures of
nonclassicality for pure Gaussian states; $\tau _{{\rm m}}^{(\zeta )}$ goes
from $0$ to $1/2$ and $d_{{\rm m}}^{(\zeta )}$ varies from $0$ to $1$ as $r$
raises from $0$ to $\infty $. This equivalence has also been demonstrated
for a Bures-Uhlmann distance-type measure \cite{Marian}; this indicates that
phase-space and distance-type measures of nonclassicality are equivalent for
the set Gaussian pure states. Actually, all measures based on distances
that, for pure states, depend only on $\left| \left\langle \psi |\varphi
\right\rangle \right| $, with coherent states elected as the most
nonclassical ones, are equivalent to the phase-space measure as far squeezed
states are concerned.

On the other hand, for non-Gaussian pure states, like the superpositions of
states analyzed before with the phase-space measure, such an equivalence
does not exist, an aspect already observed for number states. In fact, for
the general superposition of the vacuum and the one-photon state 
\begin{equation}
\left| \psi (\xi ;\phi )\right\rangle _{0,1}=\sqrt{\xi }\,\left|
0\right\rangle +\sqrt{1-\xi }\,e^{i\phi }\,\left| 1\right\rangle \;,
\label{vac11}
\end{equation}
the $Q$-function is given by (taking $\beta =be^{i\theta }$) 
\begin{equation}
Q_{\left| \psi \right\rangle _{0,1}}(\beta )=\frac{1}{\pi }e^{-b^{2}}\left[
\xi +(1-\xi )b^{2}+2\sqrt{\xi (1-\xi )}b\cos (\theta -\phi )\right] \;,
\label{Qfunc01}
\end{equation}
and the degree of nonclassicality (\ref{DTmeasu}) is found to be 
\begin{equation}
d_{{\rm m}}^{(0,1)}(\xi )=1-\exp \left[ -\frac{2-\xi -\sqrt{\xi (4-3\xi )}}{ 
2(1-\xi )}\right] \left[ 1+\frac{1}{2}\sqrt{\xi (4-3\xi )}-\frac{1}{2}\xi  
\right] \;.  \label{dm01}
\end{equation}
One sees that $d_{{\rm m}}^{(0,1)}(\xi )$ decreases monotonically from $d_{ 
{\rm m}}^{(1)}=1-e^{-1}$ (corresponding to the number state $\left|
1\right\rangle $) to $d_{{\rm m}}^{(0)}=0$ (the degree of nonclassicality of
the vacuum) as $\xi $ varies from $0$ to $1$, a behavior rather distinct
from the phase space-measure which gives $\tau _{{\rm m}}^{(0,1)}=1$ for all 
$0\leq \xi <1$. This example shows clearly that distance-type measures are
efficient to quantify the degree of nonclassicality of non-Gaussian pure
states, at least for measures based on the Hilbert-Schmidt distance, in
contrast with the phase-space measure which assigns maximum nonclassical
depth to all of them.

As another example, consider the even ($+$) and the ($-$) coherent states
(Schr\"{o}dinger-cat states with well-defined parities), with $\alpha \in 
{\bf R}^{+}$ for simplicity, 
\begin{equation}
\left| \Psi \right\rangle _{\pm }=\frac{1}{\sqrt{2\left[ 1\pm \exp (-2\alpha
^{2})\right] }}\left( \left| -\alpha \right\rangle +\left| \alpha
\right\rangle \right) \;.  \label{Sevcat}
\end{equation}
Their $Q$-functions, obtained from Eqs. (\ref{Rcat}) and (\ref{Fcat}) by
making $\tau =1$, $\xi =1/2$ and $\phi _{+}=0$ (or $\phi _{-}=\pi $), are
given by 
\begin{equation}
Q_{\pm }(x,y)=\frac{\exp \left[ -(x^{2}+y^{2})\right] }{\pi \left[ \exp
(\alpha ^{2})\pm \exp (-\alpha ^{2})\right] }\left[ \cosh (2\alpha x)\pm
\cos (2\alpha y)\right] \;.  \label{Qfevcat}
\end{equation}
For even states, maximum values of the functions $Q_{+}$, which are
certainly attained at points along the $y=0$ line, can not be analytically
calculated for arbitrary $\alpha $, but the degree of nonclassicality can be
easily evaluated by numerical means. For $0\leq \alpha \leq 1$, maxima occur
at $x=0$ and one finds exactly 
\begin{equation}
d_{{\rm m}}^{(+)}(\alpha )=1-{\rm sech}\,(\alpha ^{2})\;.  \label{dmcat01}
\end{equation}
When $\alpha $ is increased above $1$, two maxima of the function $Q_{+}$
(for each $\alpha $) exist at values of $x$ which approaches $-\alpha $ and $ 
\alpha $; for $\alpha =2.7$, maxima occurs at $|x|=2.699997$ in a six
decimal-places precision. Thus, asymptotically (for large $\alpha $) 
\begin{equation}
d_{{\rm m}}^{(+)}(\alpha )\simeq \frac{1}{2}\left[ 1-\exp (-2\alpha ^{2}) 
\right] \;,  \label{dmcatassym}
\end{equation}
and one sees that the degree of nonclassicality, relative to the measure ( 
\ref{DTmeasu}), of even coherent states grows continuously from $0$ to $1/2$
as $\alpha $ varies from $0$ to $\infty $, while $\tau _{{\rm m}}^{(+)}=1$
for all $\alpha >0$. The increasing of $d_{{\rm m}}^{(+)}(\alpha )$, meaning
that the even coherent state becomes more nonclassical as $\alpha $
augments, is in agreement with the fact that the decoherence time diminishes
as the separation (in the phase space) of the components of a superposition
of coherent states is enlarged \cite{Brune}. Similarly, investigating the
maxima of $Q_{-}(x,y)$ as $\alpha $ is changed, one finds that the degree of
nonclassicality of odd coherent states, $d_{{\rm m}}^{(-)}(\alpha )$, varies
continuously from the value $d_{{\rm m}}^{(1)}=1-e^{-1}$ (corresponding to
the number state $\left| 1\right\rangle $ for which $\left| \Psi
\right\rangle _{-}$ tends when $\alpha \rightarrow 0$) to $1/2$ as $\alpha $
is increased from $0$ to $\infty $.

The analysis of the degree of nonclassicality of mixed states using
distance-type measures is more involved since, in this case, one has to use
the expressions (\ref{HSdist}) or (\ref{BUdist}), which are much less
operational than (\ref{HSdistpu}) and (\ref{BUdistpu}), and to extend the
set of the most nonclassical states \cite{Dodonov,Marian}. This issue and
the discussion of the relation between the degree of nonclassicality and the
occurrence of nonclassical properties are left for future work.

\section{Conclusions}

The phase-space measure of nonclassicality of states of a field mode has
been discussed for both superpositions and mixtures of pure states. General
superpositions of the vacuum and number states and of opposite coherent
states were considered to show explicitly that pure non-Gaussian states
possess maximum nonclassical depth, while mixtures of representative pure
states have (in general) nonclassical depth interpolating between the values
corresponding to their added parts. Distance-type measures were also
analyzed. It was demonstrated that all measures of nonclassicality defined
as the minimum value of any monotonically increasing function of a distance
between the state and the set of all coherent states, which can be expressed
sole \ in terms of the quantum-mechanical transition probability between the
states, are equivalent to the phase-space measure when one deals with
pure-Gaussian (squeezed) states, the degree of nonclassicality being a
bijective function of the nonclassical depth. For non-Gaussian pure states,
examples were examined showing that distance-type measures are more
appropriated to quantify the degree of nonclassicality in this case. The
more involving problem of quantifying the nonclassicality of mixed states
within the viewpoint of distance-type measures will be discussed elsewhere.

\ 

{\bf Acknowledgments}

This work was partially supported by CNPq, Brazil.

\bigskip \ 

{\bf Figure caption}

\ 

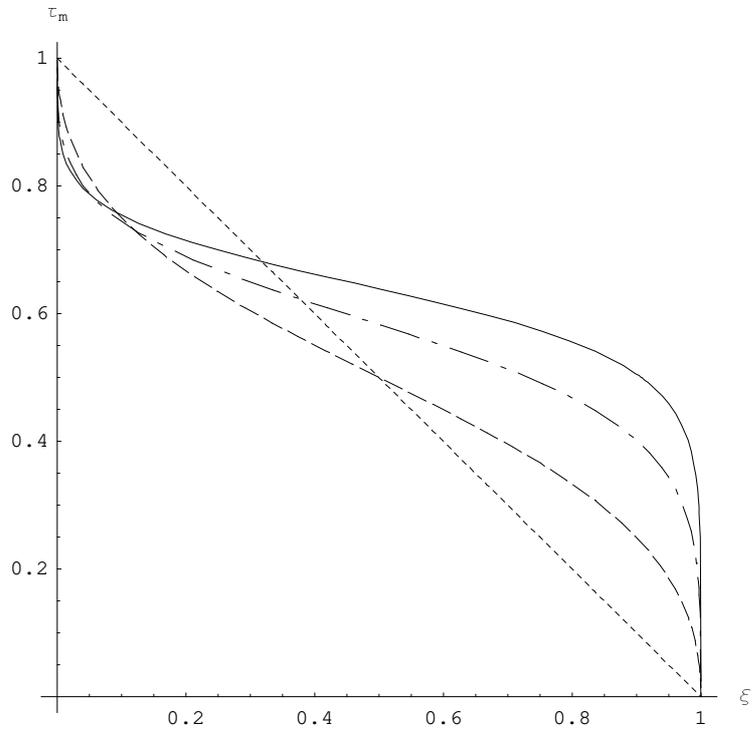
Fig. 1. Nonclassical depth ($\tau _{{\rm m}}$) versus the interpolating
parameter ($\xi $) for mixtures of the vacuum and a $n$-photons state, Eq. ( 
\ref{mix}). Dotted, dashed, dot-dashed and full lines correspond to $n=1$, $ 
2 $, $3$ and $4$, respectively.

\end{document}